\documentclass[fleqn,10pt,twocolumn]{wlscirep}
\usepackage[utf8]{inputenc}
\usepackage[T1]{fontenc}
\usepackage{bm}
\usepackage{cite}
\usepackage{float}
\usepackage{setspace}
\usepackage{microtype}
\usepackage{booktabs}
\usepackage{array}
\usepackage{caption}
\usepackage{hyperref}

\title{QHap: Quantum-Inspired Haplotype Phasing}

\author[1,3,\dag]{Rui Zhang}
\author[2,4,\dag]{Xian-Zhe Tao}
\author[3,\dag]{Yibo Chen}
\author[1,3,\dag]{Jiawei Zhang}
\author[5,\dag]{Lei He}
\author[3]{Dongming Fang}
\author[5,7]{Lin Yang}
\author[5]{Yuhui Sun}
\author[1,3]{Qinyuan Zheng}
\author[3,8]{Xinmeng Shi}
\author[3]{Yang Zhou}
\author[3,7]{Wanyi Chen}
\author[3,6,*]{Chentao Yang}
\author[2,4,*]{Man-Hong Yung}
\author[3,6,*]{Jun-Han Huang}

\affil[1]{School of Artificial Intelligence, University of Chinese Academy of Sciences, Beijing, 100049, China}
\affil[2]{International Quantum Academy, Shenzhen, 518048, China}
\affil[3]{State Key Laboratory of Genome and Multi-omics Technologies, BGI Research, Shenzhen, 518083, China}
\affil[4]{Shenzhen Institute for Quantum Science and Engineering, Southern University of Science and Technology, Shenzhen, 518055, China}
\affil[5]{MGI Tech, Shenzhen, 518083, China}
\affil[6]{Shenzhen Key Laboratory of Transomics Biotechnologies, BGI Research, Shenzhen, 518083, China}
\affil[7]{College of Life Sciences, University of Chinese Academy of Sciences, Beijing, 101408, China}
\affil[8]{College of Life Sciences, Northwest University, Xi'an, 710069, China}
\affil[$\dag$]{These authors contributed equally to this work}
\affil[*]{Correspondence: yangchentao@genomics.cn (C.Y.), yung@iqasz.cn (M.-H.Y.), huangjunhan@genomics.cn (J.-H.H.)}

\begin{abstract}
Haplotype phasing, the process of resolving parental allele inheritance patterns in diploid genomes, is critical for precision medicine and population genetics, yet the underlying optimization is NP-hard, posing a scalability challenge. To address this, we introduce QHap, a haplotype phasing algorithm that leverages quantum-annealing-inspired optimization. By reformulating haplotype phasing as a Max-Cut problem and deploying a GPU-accelerated ballistic simulated bifurcation solver, QHap accelerates phasing while maintaining accuracy comparable to established phasing tools. On the highly polymorphic human major histocompatibility complex region, QHap demonstrates 4- to 20-fold acceleration over HapCUT2 and WhatsHap with zero switch error across multiple long-read sequencing platforms. The framework implements two strategies: a read-based method for regional phasing, and a single nucleotide polymorphism-based method that, through quality-weighted probabilistic edge construction, efficiently scales to chromosome-scale tasks. Integration of Pore-C chromatin conformation capture data increases the haplotype N50 by up to 15-fold, enabling near-chromosome-scale haplotype reconstruction. QHap demonstrates that quantum-inspired algorithms operating on classical hardware offer a promising approach to addressing the growing computational demands of sequencing data, establishing a new paradigm for applying physics-inspired optimization to fundamental challenges in computational genomics.
\end{abstract}
\begin{document}

\flushbottom
\maketitle

\thispagestyle{empty}

\section*{Introduction}

Haplotype phasing is the process of distinguishing the alleles inherited together on a chromosome from a parent in diploid genomes. Haplotypes are essential for interpreting the genetic mechanisms underlying biological phenotypes and for advancing haplotype-resolved genome assembly. In human genetics, phased haplotypes enable precise mapping of compound heterozygosity and allele-specific expression, which are critical for diagnosing complex diseases such as cystic fibrosis and $\beta$-thalassemia, as well as for reconstructing population migration histories \cite{r1, r2, r3, r4, huang2026snp, cao2026cutehap}.
 
In recent years, long-read sequencing technologies have significantly advanced haplotype phasing by producing reads spanning 10 to 100 kilobases (kb) \cite{r5, r6}. PacBio High-Fidelity (HiFi) and Oxford Nanopore Technologies (ONT) are widely adopted platforms that combine extended read lengths with high single-base accuracy. CycloneSEQ, a recently developed nanopore-based platform, offers enhanced throughput for cost-effective genome-scale applications \cite{r7, r8}. Extended reads capture multiple variant sites per fragment, facilitating genome-wide haplotype reconstruction without reliance on external references or familial data \cite{r9, r10}.

Despite these advances, sequencing and alignment errors introduce genotype inaccuracies at single nucleotide polymorphism (SNP) sites, compounding the inherent computational complexity of haplotype phasing—a problem proven to be NP-hard \cite{r11}. Existing long-read phasing tools address this challenge through different optimization strategies. HapCUT2 iteratively refines haplotypes using likelihood-based graph cuts, whereas WhatsHap solves the weighted minimum error correction (wMEC) problem by dynamic programming. Although these methods achieve high accuracy, their computational demands escalate with read length and dataset size, and current implementations lack GPU support, hindering population-scale analyses amid declining sequencing costs and expanding precision medicine initiatives \cite{r13, r14, r15, r16, r17, EDVQE}.

To address this limitation, we introduce QHap, which reformulates haplotype phasing as a Max-Cut problem and solves it using ballistic simulated bifurcation (bSB), a quantum-inspired optimization algorithm \cite{r14}. By simulating a nonlinear Hamiltonian system in which continuous spin variables evolve under ballistic dynamics on a gradually deformed energy landscape, bSB enables efficient exploration of large combinatorial solution spaces and is naturally suited to GPU parallelism \cite{r18,r19,r20,r21,r22}. In this framework, the graph partition directly yields the phase assignment rather than serving only as an intermediate refinement step. QHap implements this idea in two forms: a read-based formulation that partitions sequencing fragments according to allelic discordance, and an SNP-based formulation that partitions variant loci using aggregated quality-weighted evidence for \textit{cis} versus \textit{trans} relationships. Using this strategy, the GPU-accelerated bSB solver achieves major histocompatibility complex (MHC) region phasing in approximately 1 minute, over an order of magnitude faster than HapCUT2 and WhatsHap running on a single CPU thread. Furthermore, QHap supports the integration of multi-way contacts from Pore-C chromatin conformation capture data, which enhances long-range connectivity and boosts haplotype N50 values, facilitating chromosome-scale haplotype reconstruction \cite{r31, r24}.

\section*{Results}
\subsection*{Overview of QHap}

\begin{figure*}[h!]
    \centering
    \includegraphics[width=0.97\textwidth,keepaspectratio]{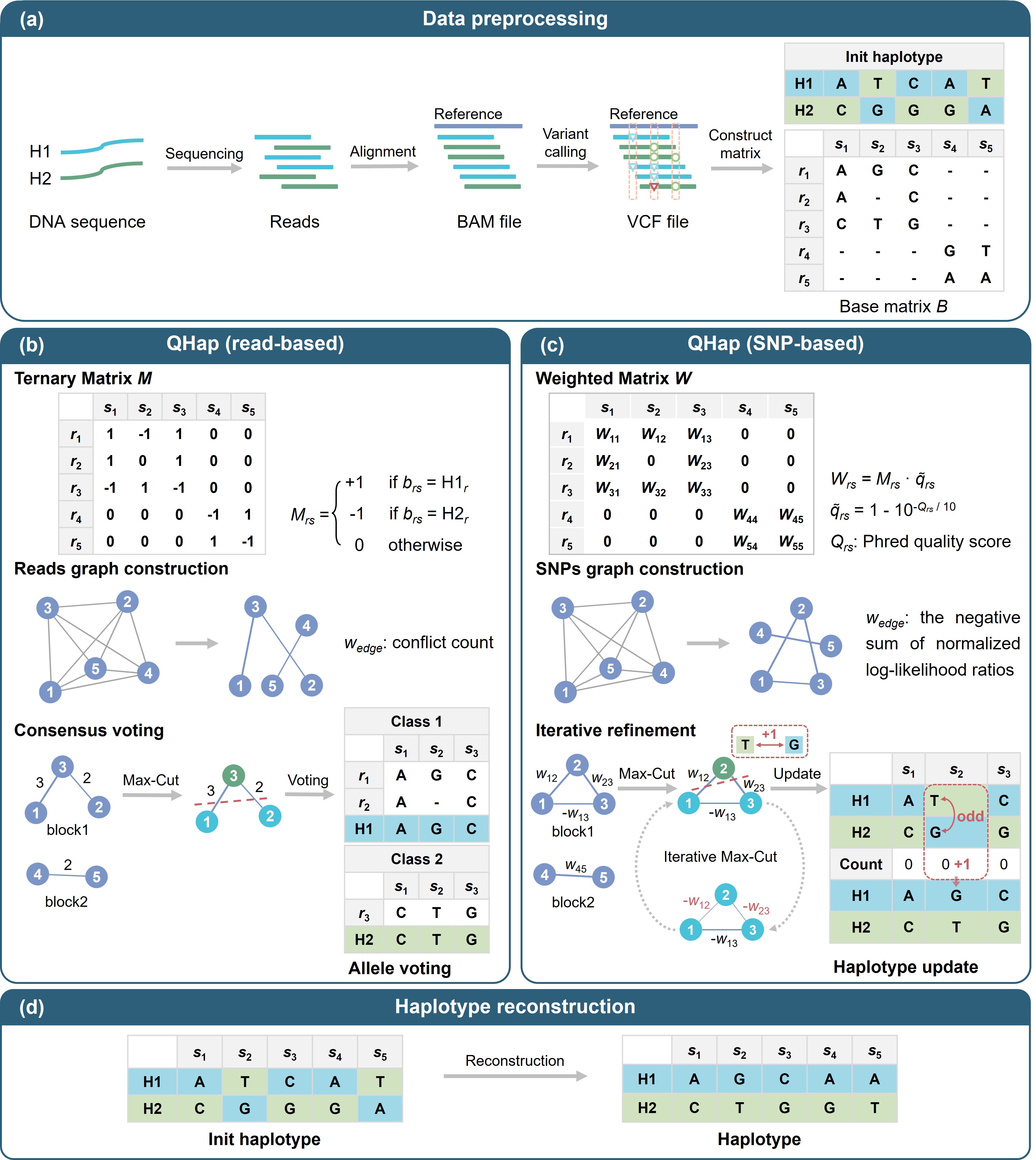}
    \caption{\textbf{Overview of the QHap framework.}
    QHap takes aligned reads (BAM) and variant calls (VCF) as input,
    constructs a read--SNP base matrix~$B$, and initializes a haplotype
    pair during preprocessing~(\textbf{a}).
    In the read-based method~(\textbf{b}), $B$ is encoded into a ternary
    matrix~$M$ from which a weighted graph is built with reads as
    vertices, where each edge weight counts the number of shared loci
    carrying opposing alleles. Connected components are partitioned via
    Max-Cut into phasing blocks, and consensus haplotypes are resolved by
    majority allele voting.
    In the SNP-based method~(\textbf{c}), a quality-weighted matrix~$W$
    combines the ternary encoding with Phred-derived error probabilities.
    A weighted graph is then constructed with SNPs as vertices, where
    edge weights are defined as the negative sum of normalized
    log-likelihood ratios. After connected components are partitioned via
    Max-Cut, non-trivial partitions undergo iterative vertex
    reassignment with sign-flipped weights until convergence, and the
    accumulated per-locus node counts trigger H1/H2 allele swaps at
    positions with odd counts.
    Both strategies converge on a final reconstruction step~(\textbf{d}):
    the read-based method integrates block-wise voting results, whereas
    the SNP-based method applies position-specific swap decisions to the
    initial haplotype to yield the final phased haplotypes.}
    \label{fig:1}
\end{figure*}

QHap implements two complementary graph-based strategies for haplotype phasing (Figure~\ref{fig:1}), each suited to different application scenarios. Both strategies share a common data preprocessing stage (Figure~\ref{fig:1}a) that transforms raw sequencing alignments and variant calls into a structured base matrix, from which an initial haplotype pair is established for subsequent allelic encoding.

From this foundation, the two methods diverge in how they formulate the phasing problem as a graph optimization task. The read-based method (Figure~\ref{fig:1}b) constructs a graph in which
sequencing reads serve as vertices and each edge weight counts the
number of shared SNP positions at which two reads carry opposing
alleles. This approach directly captures read-to-read relationships for precise local phase resolution. Yet its complexity scales with the number of reads, making it more suitable for targeted regional analysis or moderate-coverage datasets. However, as sequencing technologies advance toward megabase-scale read lengths, individual fragments may span entire haplotype blocks, positioning the read-based method for direct phase resolution from complete fragment information without requiring inference across SNP linkages.

In contrast, the SNP-based method (Figure~\ref{fig:1}c) constructs a
graph in which variant loci serve as vertices and edge weights are defined as the negative sum of normalized log-likelihood ratios quantifying the evidence for \textit{cis} versus \textit{trans} haplotype relationships. Because graph size depends on the number of variants rather than sequencing depth, higher sequencing depth enhances edge reliability via accumulated evidence without expanding the graph size, thereby preserving computational tractability for chromosome-scale phasing. By transforming the search for the optimal Max-Cut solution into an iterative refinement procedure, this method further improves convergence on the complex optimization landscape arising from mixed-sign edge weights. Both methods converge on a final haplotype reconstruction step (Figure~\ref{fig:1}d), which either directly generates phased haplotypes by integrating block-wise voting results, or updates the initialized haplotypes via position-level allele swaps guided by partition statistics to yield the phased haplotypes.

\subsection*{Performance}
To evaluate the performance of QHap, we benchmark both the read-based and SNP-based methods against HapCUT2 and WhatsHap, two widely used haplotype phasing tools, using the Genome in a Bottle (GIAB) HG002 sample \cite{r25}. All experiments were conducted on a system equipped with an Intel Core i7-14650HX processor (2.20 GHz) and an NVIDIA GeForce RTX 5060 Laptop GPU. The MHC region on chromosome 6 is selected as the benchmark region due to its clinical significance in transplantation compatibility and autoimmune disease susceptibility, as well as its complex genomic architecture characterized by extensive structural variation and segmental duplications that pose challenges for accurate read alignment \cite{r26}. 

We assess phasing quality using six metrics: switch error rate (SE), hamming error rate (HE), haplotype N50, phasing completeness, SNP completeness, and running time (detailed definitions in Methods). As shown in Table \ref{table:1}, across all three long-read sequencing platforms (CycloneSEQ, HiFi, and ONT) at 50× coverage depth, QHap achieves phasing accuracy comparable to established methods. Both QHap and HapCUT2 report zero SE and HE on all datasets, while WhatsHap exhibits marginally elevated error rates (0.02\% for both SE and HE) on CycloneSEQ data. Notably, all tools achieve nearly identical haplotype N50 values, phasing completeness, and SNP completeness within each platform, demonstrating that QHap maintains accuracy parity with HapCUT2 and WhatsHap.

The primary advantage of QHap lies in its computational efficiency. QHap leverages a GPU-accelerated bSB solver to achieve considerable speedups compared to HapCUT2 and WhatsHap, which rely on single-threaded CPU execution. On CycloneSEQ data, QHap completes phasing in 63 to 70 seconds, representing a 4.4-fold speedup over WhatsHap (304.1 s) and a 14.4-fold speedup over HapCUT2 (913.0 s). This efficiency gain is consistent across platforms: on HiFi data, the SNP-based method achieves a 19-fold speedup over HapCUT2, while on ONT data, QHap demonstrates an 8.5- to 20-fold improvement. Such time reductions are valuable for large-scale population studies requiring high aggregate throughput and for clinical applications demanding rapid sample processing.

A noteworthy observation emerges from cross-platform comparison of haplotype contiguity. CycloneSEQ achieves the highest N50 value (3,830.5 kb) compared with HiFi (583.6 kb) and ONT (805.9 kb), indicating superior phase block contiguity that likely benefits from its longer read lengths enabling broader variant bridging. Comparing the two QHap strategies across these platforms reveals distinct computational trade-offs. The SNP-based method demonstrates faster execution on the HiFi and ONT datasets, where graph complexity scales with variant count rather than read number. In contrast, the read-based method shows competitive runtime on CycloneSEQ data but incurs higher computational costs on the other two platforms, consistent with its graph size scaling with the number of sequencing fragments. This observation suggests that the read-based method may be particularly suited for emerging sequencing technologies that prioritize read length, as it can effectively leverage extended genomic spans for haplotype resolution even in the presence of moderate sequencing errors.

\begin{table*}[h!]
\centering
\captionsetup{justification=raggedright, singlelinecheck=false}
\caption{Performance of QHap and other haplotype phasing tools on different datasets in the MHC region}
\resizebox{\linewidth}{!}{
    \begin{tabular}{llcccccc}
    \toprule
    Dataset & Tool & SE (\%) & HE (\%) & Haplotype & Phasing & SNP completeness & Running \\
    (50$\times$ coverage) &&&& N50 (kb)& completeness (\%) &(\%) & time (s) \\
    \midrule
    HG002      & QHap (read-based) & 0.00 & 0.00    & 3830.545 & 100.00 & 97.73 & \textbf{63.454}\\
    CycloneSEQ & QHap (SNP-based) & 0.00 & 0.00    & 3830.545 & 100.00   & 97.73 & 69.634   \\
               & HapCUT2          & 0.00 & 0.00    & 3830.545 & 100.00   & 97.73 & 913.032  \\
               & WhatsHap         & 0.02 & 0.02 & 3830.545 & 100.00 & 97.73 & 304.100    \\
    \midrule
    HG002      & QHap (read-based) & 0.00 & 0.00    & 583.591& 99.98 & 99.64 & 89.178   \\
    HiFi       & QHap (SNP-based) & 0.00 & 0.00    & 583.591  & 99.98 & 99.64 & \textbf{45.278}   \\
               & HapCUT2          & 0.00 & 0.00    & 583.591  & 99.98 & 99.64 & 858.700    \\
               & WhatsHap         & 0.00 & 0.00    & 583.591  & 99.95 & 99.62 & 318.600    \\
    \midrule
    HG002      & QHap (read-based) & 0.00 & 0.00    & 805.858  & 99.99 & 98.17 & 115.709  \\
    ONT        & QHap (SNP-based) & 0.00 & 0.00    & 805.858  & 99.99 & 98.17 & \textbf{57.059}   \\
               & HapCUT2          & 0.00 & 0.00    & 805.858  & 99.99 & 98.17 & 1150.967 \\
               & WhatsHap         & 0.00 & 0.00    & 805.858  & 99.99 & 98.17 & 493.300    \\
    \bottomrule
    \end{tabular}
}
\par
\vspace*{2pt}
\footnotesize 
\parbox{\linewidth}{SE: switch error rate; HE: hamming error rate. Running time represents the complete pipeline including data preprocessing and phasing. All experiments were performed on a system with an Intel(R) Core(TM) i7-14650HX processor (2.20~GHz) and an NVIDIA GeForce RTX 5060 Laptop GPU. HapCUT2 and WhatsHap were executed on a single CPU thread. QHap utilized 8 CPU threads for preprocessing and GPU-accelerated bSB for Max-Cut optimization. For the read-based method, bSB parameters were set to 10,000 iterations with 200 samples for CycloneSEQ and HiFi data, and 20,000 iterations with 1,000 samples for ONT data. For the SNP-based method, bSB parameters were adaptively configured based on graph size (500 iterations for graphs with <5,000 nodes; 1,000 iterations otherwise; both with 100 samples), with a maximum convergence count of 15.}
\label{table:1}
\end{table*}

\subsection*{Benchmarking Max-Cut algorithms}

To evaluate the suitability of bSB for Max-Cut optimization in QHap, we systematically benchmark six algorithms on graphs constructed from CycloneSEQ long-read sequencing data of the HG002 sample in the MHC region. These include five quantum-inspired algorithms: bSB, discrete simulated bifurcation (dSB), adiabatic simulated bifurcation (aSB), negative-mass field annealing (NMFA), and simulated coherent Ising machine (SimCIM), as well as classical simulated annealing (SA).

\begin{figure*}[h!]
    \centering
    \includegraphics[width=0.95\textwidth,keepaspectratio]{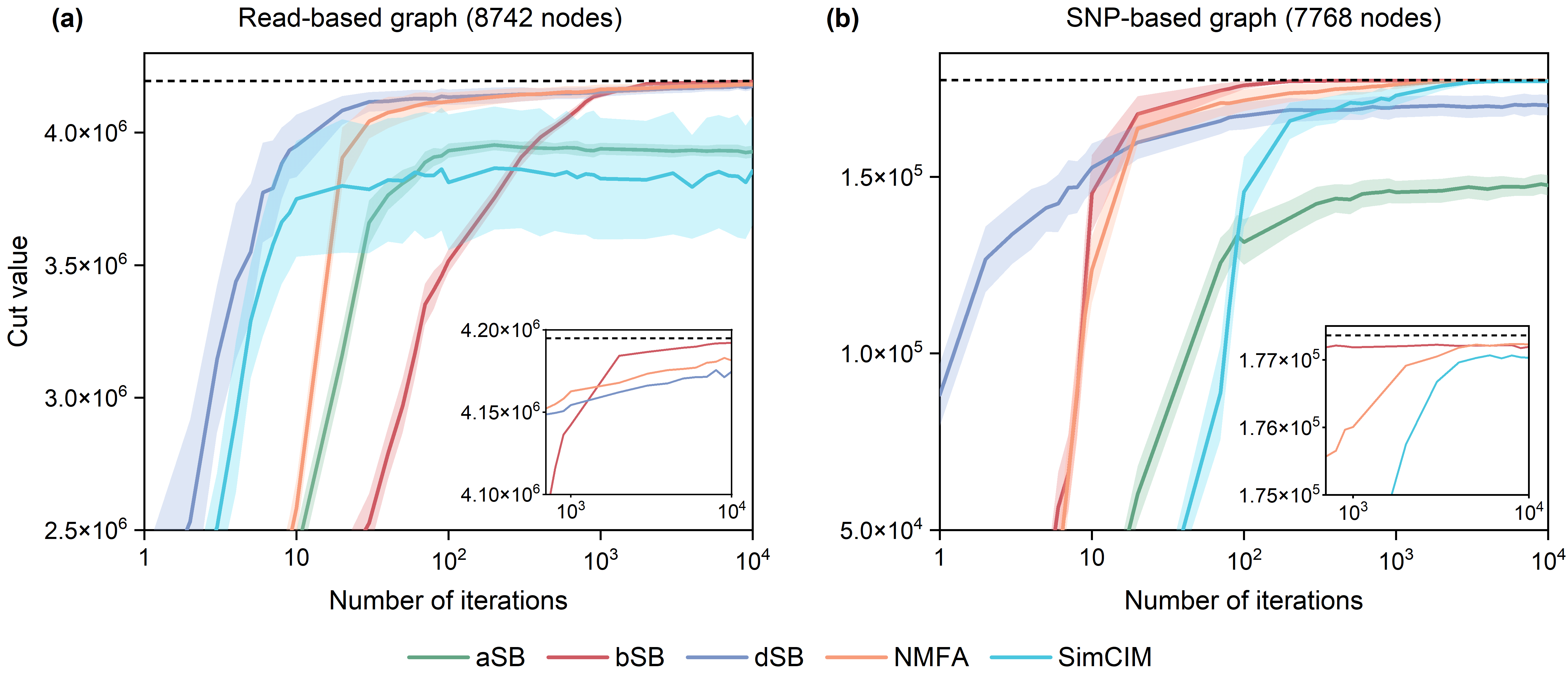}
    \caption{\textbf{Performance comparison of quantum-inspired
    optimization algorithms on QHap-constructed graphs.}
    Five algorithms are benchmarked on two representative connected
    components: a non-negative integer-weighted graph (8,742 nodes,
    259,417 edges) derived from the read-based formulation~(\textbf{a})
    and a mixed-sign floating-point-weighted graph (7,768 nodes,
    3,303,740 edges) derived from the SNP-based
    formulation~(\textbf{b}). Solid lines represent mean cut values
    averaged over 100 independent runs, shaded regions indicate standard
    deviation, and dashed lines mark the best value achieved during
    runtime. All algorithms were executed on an Intel Core i7-14650HX
    processor (2.20~GHz).}
    \label{fig:2}
\end{figure*}

QHap's two graph construction strategies yield distinct edge weight characteristics. The read-based method produces non-negative integer-valued weights derived from direct allelic conflict counts, while the SNP-based method generates mixed-sign floating-point weights based on quality-adjusted log-likelihood ratios. These differences in weight distribution and sign structure affect algorithm convergence behavior and solution stability. As illustrated in Figure~\ref{fig:2}, we evaluate solution quality of the five quantum-inspired algorithms on representative connected components from both formulations: panel (a) corresponds to a non-negative integer-weighted graph from the read-based method, while panel (b) corresponds to a mixed-sign floating-point-weighted graph from the SNP-based method. When the number of iterations ($n_{\mathrm{iter}}$) is sufficiently large, bSB consistently achieves favorable solution accuracy and stability on both graph types, regardless of edge weight type or graph structure.

\begin{figure*}[h!]
    \centering
    \includegraphics[width=0.97\textwidth,keepaspectratio]{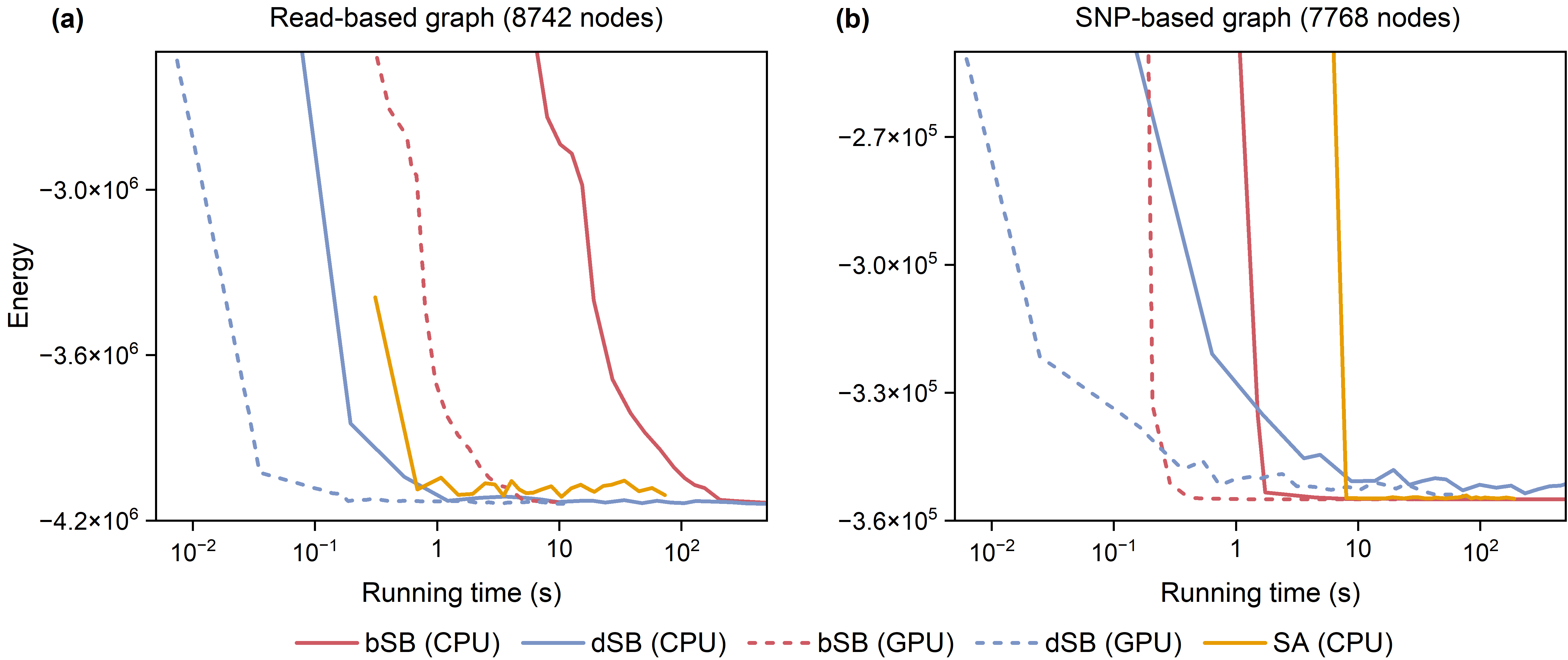}
    \caption{\textbf{Comparison of Ising energy convergence for
    quantum-inspired optimization algorithms and classical SA.}
    Energy evolution is shown for a non-negative integer-weighted graph
    (8,742 nodes, 259,417 edges) from the read-based
    formulation~(\textbf{a}) and a mixed-sign floating-point-weighted
    graph (7,768 nodes, 3,303,740 edges) from the SNP-based
    formulation~(\textbf{b}). At each time step, the plotted energy
    corresponds to the minimum across 200 parallel samples.
    GPU-accelerated algorithms were executed on an NVIDIA GeForce RTX
    5060 Laptop GPU; CPU-based execution used an Intel Core i7-14650HX
    processor (2.20~GHz).}
    \label{fig:3}
\end{figure*}

Beyond solution quality, computational efficiency is critical for practical deployment. Simulated bifurcation algorithms are particularly well-suited for GPU acceleration due to their inherently parallel spin dynamics. Figure~\ref{fig:3} demonstrates order-of-magnitude improvements in convergence speed under GPU acceleration compared to CPU-only execution and SA. However, a combined analysis of convergence speed (Figure~\ref{fig:3}) and stability (Figure~\ref{fig:2}) reveals critical trade-offs between dSB and bSB. While panel (a) in Figure~\ref{fig:3} shows that dSB exhibits faster initial convergence than bSB on the integer-weighted graph, its solutions display a larger standard deviation across independent runs (Figure~\ref{fig:2}a), indicating reduced stability. Furthermore, on the mixed-sign floating-point-weighted graph, panel (b) in Figure~\ref{fig:3} reveals that dSB exhibits erratic convergence behavior. In contrast, bSB maintains consistent and reliable convergence characteristics across both graph types. This stability likely arises from bSB's continuous ballistic dynamics: the momentum term provides inertia that enables spins to traverse shallow local minima, while the gradual evolution naturally accommodates the complex energy landscape induced by mixed-sign edge weights and continuous weight values; by contrast, dSB's discrete sign-based updates can overshoot optimal configurations on such landscapes, where competing positive and negative edges create frustrated regions that require smooth navigation rather than abrupt sign flips\cite{r20, r33, goto2026edge, tao2026tabu}.

Collectively, bSB demonstrates robust adaptability across diverse edge weight types and graph structures. It exhibits a smaller standard deviation in converged solutions across multiple runs, indicating more consistent results from independent executions. Moreover, bSB enables efficient GPU acceleration, yielding order-of-magnitude speedups over CPU-only execution. This combination of solution quality and computational efficiency establishes bSB as the appropriate algorithm for the QHap framework.

\subsection*{SNP linkage depth analysis}

The SNP-based graph construction employs a linkage depth parameter (ranging from 1/10 to 10/10) that controls the extent of pairwise SNP connections within each read. Using CycloneSEQ data of the HG002 sample in the MHC region as a case, this section analyzes the trade-off between computational cost and phasing accuracy as governed by this parameter.

\begin{figure*}[h!]
    \centering
    \includegraphics[width=0.8\linewidth,keepaspectratio]{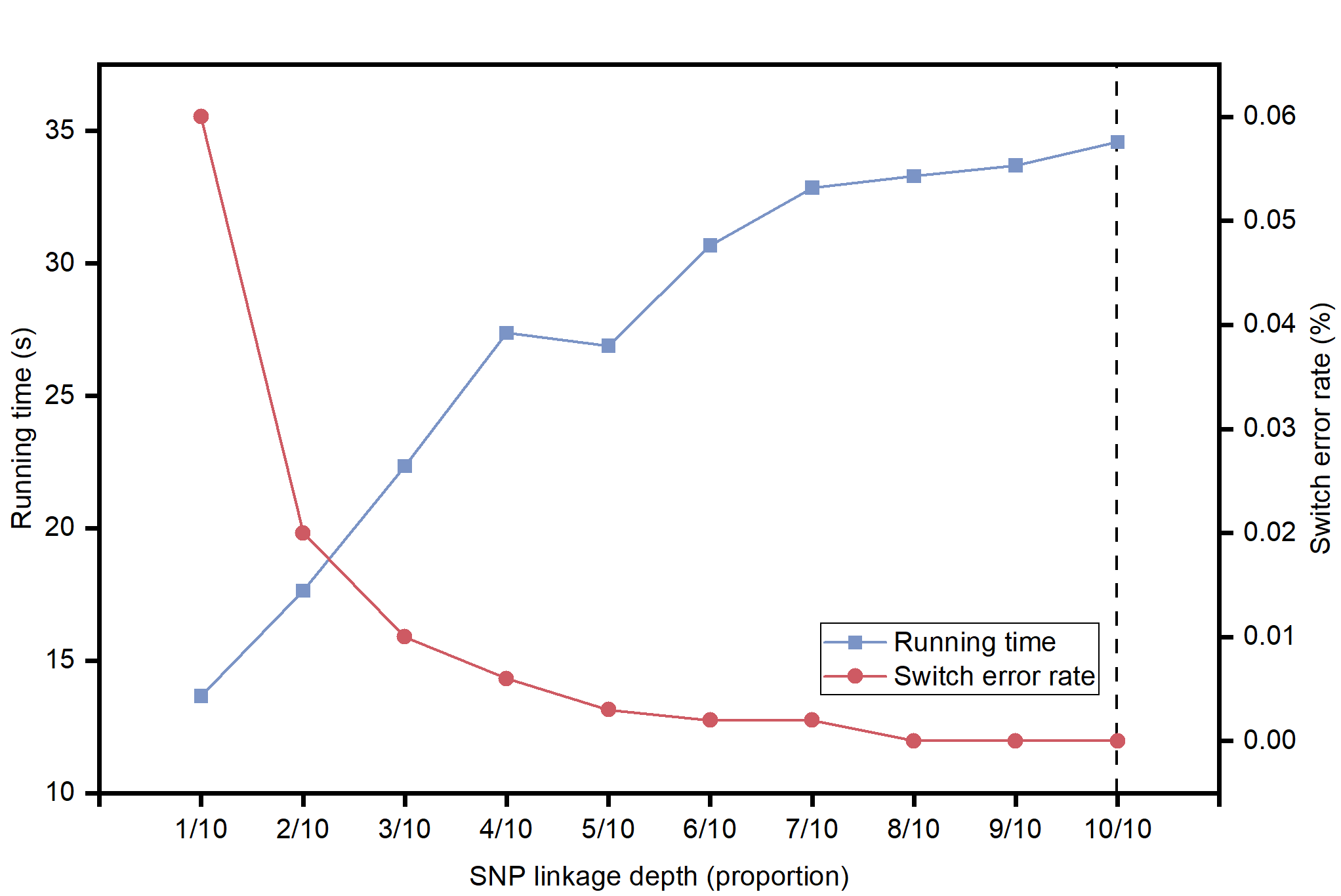}
    \caption{\textbf{Impact of SNP linkage depth on phasing accuracy and
    computational cost.}
    The linkage depth proportion (x-axis, ranging from 1/10 to 10/10)
    controls the fraction of downstream SNP loci with which each SNP
    forms pairwise connections during graph construction. The left y-axis
    shows running time of the core phasing algorithm (excluding data
    loading and preprocessing), and the right y-axis shows the switch
    error rate (SE). The vertical dashed line indicates the linkage depth parameter 
    adopted in this paper. The bSB parameters were adaptively configured 
    based on graph size (500 iterations for graphs with < 5,000 nodes and 
    1,000 iterations otherwise, both utilizing 100 samples), with a 
    maximum convergence count of 5. All measurements represent the mean 
    values obtained from 10 independent runs of the algorithm. Experiments 
    were performed on a system with an Intel(R) Core(TM) i7-14650HX 
    processor (2.20~GHz) and an NVIDIA GeForce RTX 5060 Laptop GPU.}
    \label{fig:4}
\end{figure*}

Figure \ref{fig:4} illustrates this trade-off. If a read covers $N$ heterozygous SNP loci ordered by genomic position, a linkage depth of $k/10$ means that each SNP is connected only to its nearest $\lfloor k \times (N-1) / 10 \rfloor$ downstream SNPs within that read. This deterministic selection prioritizes proximal linkages, which typically carry stronger phase evidence. At the extreme of exhaustive linkage (10/10), every possible SNP pair within each read is examined, capturing all available linkage information. This comprehensive approach achieves perfect phasing accuracy with zero switch error, as all transitive phase relationships are detected and reinforcing evidence from multiple linkage paths resolves ambiguities. However, exhaustive examination incurs a computational cost (34.58 s; in this section, running time refers exclusively to the core phasing algorithm execution, excluding data loading and preprocessing) due to the quadratic growth in edge evaluations.

As the linkage depth decreases, runtime improves at the expense of phasing accuracy. At an 8/10 depth, the SE remains at 0\% while the runtime decreases, indicating that some edge redundancy exists in the complete graph. A further reduction to 6/10 introduces minimal SE (0.002\%) with continued runtime improvement. At the minimal depth (1/10), the algorithm achieves the fastest execution (13.66 s) but results in an elevated SE (0.06\%), as critical long-range dependencies between distant SNPs remain undetected, leading to erroneous haplotype phase assignments.

These results suggest that an intermediate linkage depth (6/10 to 8/10) offers a favorable balance between phasing fidelity and computational tractability, which is particularly relevant for extending QHap to whole-genome scale analyses. In all relevant experiments in this paper, we adopt exhaustive linkage (10/10) to fully utilize all pairwise phase evidence within each read, thereby ensuring phasing accuracy.

\subsection*{Chromosome-scale phasing}

To assess QHap's scalability beyond localized regions, we extend benchmarking to chromosome 22 (chr22), which spans a substantially longer genomic region with a greater number of heterozygous variants compared to the MHC region \cite{r27}. The results are detailed in Table \ref{table:2}. Since the read-based method constructs graphs where vertices correspond to sequencing fragments, the resulting graph at the chromosome scale contains tens of thousands of nodes, approaching the current capacity constraints of quantum-inspired optimization algorithms. We therefore employ the SNP-based method for chromosome-scale analysis in the current implementation, as its computational complexity scales with the variant number rather than the fragment number.

Regarding computational runtime, QHap achieves 3- to 7-fold speedups across all platforms at the chromosome scale. On CycloneSEQ data, QHap completes phasing in 57.4 seconds versus 312.1 seconds for HapCUT2 and 176.4 seconds for WhatsHap. Similar runtime improvements are observed for the ONT and HiFi datasets, demonstrating that the computational advantages of QHap's GPU-accelerated framework extend robustly to larger genomic scales.

In terms of phasing contiguity, CycloneSEQ yields an N50 of 18,342 kb at the chromosome level, approximately 27-fold and 12-fold longer than HiFi (684.4 kb) and ONT (1,548.5 kb). This contiguity enables near-complete chromosome-scale haplotype blocks, which are essential for analyzing long-range allelic interactions and structural variant phasing.

Concerning phasing accuracy, QHap achieves SE and HE comparable to HapCUT2 and WhatsHap on HiFi and ONT data. However, the CycloneSEQ data reveals a current limitation: while SE remains minimal (0.01\%), the HE increases to 24.04\%. This elevated hamming error is primarily attributable to long switch errors, specifically isolated phase inversions affecting extended genomic segments, rather than distributed single-site misassignments. Such errors arise because the expanded solution space and increased graph complexity at the chromosome scale challenge the bSB algorithm's ability to converge to global optima within practical iteration limits, resulting in suboptimal partitions that manifest as long-range phase inversions. Notably, the near-zero short-range switch error indicates that local haplotype structure remains correctly resolved. Targeted algorithmic refinements to enhance global convergence represent a priority for future development \cite{r28}.

\begin{table*}[h!]
\centering
\captionsetup{justification=raggedright, singlelinecheck=false}
\caption{Performance of QHap and other haplotype phasing tools across different datasets on chr22}
\resizebox{\linewidth}{!}{
    \begin{tabular}{llcccccc}
    \toprule
    Dataset & Tool & SE (\%) & HE (\%) & Haplotype & Phasing & SNP completeness & Running \\
    (50$\times$ coverage) &&&& N50 (kb)& completeness (\%) &(\%) & time (s) \\
    \midrule
    HG002      & QHap (SNP-based) & 0.01 & 24.04  & 18342.065 & 100.00  & 96.42 & \textbf{57.352} \\
    CycloneSEQ & HapCUT2         & 0.00 & 0.00   & 18342.065 & 100.00  & 96.42 & 312.129 \\
               & WhatsHap        & 0.00 & 0.00   & 18342.065 & 99.99   & 96.42 & 176.400 \\
    \midrule
    HG002      & QHap (SNP-based) & 0.04 & 0.39   & 684.430   & 99.91   & 98.82 & \textbf{82.257} \\
    HiFi       & HapCUT2         & 0.03 & 0.51   & 684.430   & 99.91   & 98.82 & 255.686 \\
               & WhatsHap        & 0.05 & 0.53   & 684.430   & 99.89   & 98.80 & 208.400 \\
    \midrule
    HG002      & QHap (SNP-based) & 0.01 & 0.89   & 1548.530  & 99.98   & 97.03 & \textbf{54.026} \\
    ONT        & HapCUT2         & 0.01 & 0.31   & 1548.530  & 99.98   & 97.03 & 392.473 \\
               & WhatsHap        & 0.03 & 0.33   & 1472.058  & 99.95   & 97.00 & 263.700 \\
    \bottomrule
    \end{tabular}
}
\par
\vspace*{2pt}
\footnotesize 
\parbox{\linewidth}{SE: switch error rate; HE: hamming error rate. Running time represents the complete pipeline including data preprocessing and phasing. All experiments were performed on a system with an Intel(R) Core(TM) i7-14650HX processor (2.20~GHz) and an NVIDIA GeForce RTX 5060 Laptop GPU. HapCUT2 and WhatsHap were executed on a single CPU thread. QHap utilized 8 CPU threads for preprocessing and GPU-accelerated bSB for Max-Cut optimization. For the SNP-based method, bSB parameters were adaptively configured based on graph size (500 iterations for graphs with <5,000 nodes; 1,000 iterations otherwise; both with 100 samples), with a maximum convergence count of 15.}
\label{table:2}
\end{table*}

\subsection*{Phasing with Pore-C data integration}

Long-range chromatin conformation data provides complementary phasing information that bridges distant phase blocks separated by homozygous regions or assembly gaps. Pore-C, an Oxford Nanopore-based chromatin conformation capture method, generates multi-way contact information from ultra-long concatemers, enabling chromosome-scale haplotype connectivity that conventional pairwise Hi-C cannot achieve \cite{r29, r30, r31, r32}. Given these advantages, integrating Pore-C data into the QHap framework represents a valuable approach to extending phasing contiguity.

\begin{figure*}[h!]
    \centering
    \includegraphics[width=0.9\textwidth,keepaspectratio]{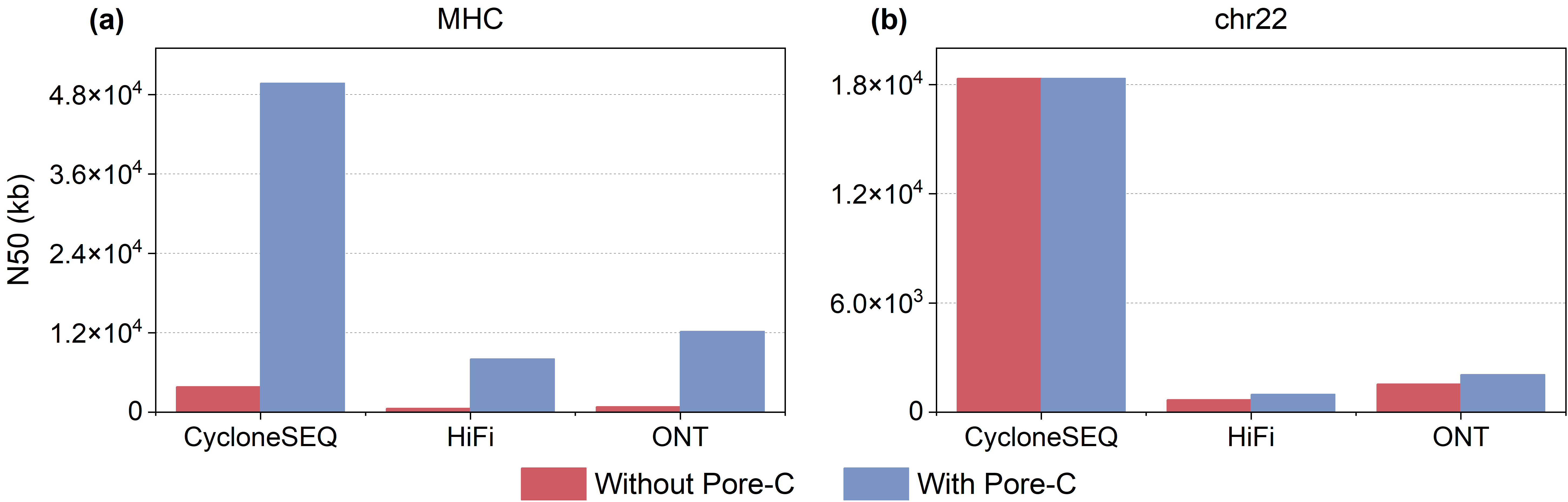}
    \caption{\textbf{Performance of QHap integrated with Pore-C data.}
    Haplotype N50 values (in kb) are compared across three sequencing
    platforms (CycloneSEQ, HiFi and ONT) with (blue) and without (red)
    Pore-C data integration for the MHC region~(\textbf{a}) and
    chr22~(\textbf{b}).}
    \label{fig:5}
\end{figure*}

QHap incorporates Pore-C data through a unified graph construction strategy: fragments derived from both long-read and Pore-C sequencing are merged based on shared read identifiers, with Pore-C contributions down-weighted to balance long-range connectivity against elevated error rates (see Methods for details). This integration substantially improves the haplotype N50 across different genomic regions and sequencing platforms, as shown in Figure \ref{fig:5}. In the MHC region, characterized by extreme polymorphism and complex haplotype structure, Pore-C integration yields marked N50 increases: phasing with CycloneSEQ data improves from 3,831 kb to 49,689 kb (13-fold), phasing with HiFi data improves from 584 kb to 8,059 kb (13.8-fold), and phasing with ONT data from 806 kb to 12,228 kb (15.2-fold). For chr22, phasing with CycloneSEQ data already yields a substantially high haplotype N50 of 18,342 kb without Pore-C integration, leaving limited headroom for further improvement; accordingly, no additional gain was observed following Pore-C integration in this case. In contrast, phasing with HiFi and ONT data exhibits substantial gains of 43.8\% (684 kb to 984 kb) and 33.3\% (1,549 kb to 2,065 kb).

Importantly, these gains in phasing contiguity are achieved without compromising phasing accuracy. Across all tested datasets, the SE and HE remain virtually unchanged following Pore-C integration, indicating that QHap's framework effectively leverages long-range contact information while preserving robust local phase inference.

Collectively, these results demonstrate that QHap can construct chromosome-scale phase blocks by integrating Pore-C data. Particularly in the highly polymorphic MHC region, this capability may extend the applicability of long-read phasing for studies requiring contiguous haplotype information, such as human leukocyte antigen (HLA) typing and structural variant characterization.

\subsection*{Proof-of-concept of QHap for HLA typing}

Despite the indispensable role of high-precision HLA typing in transplantation matching, pharmacogenomic profiling, and disease risk stratification, it remains notoriously difficult owing to the high degree of polymorphism inherent to this genomic region \cite{glasenapp2026hla}. To demonstrate the potential of QHap in facilitating downstream analyses, we applied it to HLA allele identification using long-read sequencing data from multiple platforms. The HLA genotype of the HG002 sample was obtained from a previously published high-confidence dataset and used as the benchmark.

Sequencing data generated from CycloneSEQ, HiFi and ONT were analyzed across classical HLA loci, including class I genes (HLA-A, HLA-B and HLA-C) and class II genes (DPA1, DPB1, DQA1, DQB1, DRB1 and DRB4). The inferred alleles were compared with the benchmark genotype, and the detailed results are summarized in Table \ref{table:3}.

Across most loci, the inferred alleles were consistent with the benchmark at the two-field resolution, demonstrating that the haplotype-resolved variant information produced by QHap can effectively support HLA allele identification. Accurate allele assignments were obtained for both class I and class II genes despite their high polymorphism and sequence similarity.

A small number of ambiguous calls were observed in the CycloneSEQ dataset (e.g., HLA-A and HLA-DQB1), where multiple closely related alleles were reported. This likely reflects the high sequence similarity among alleles in the IMGT/HLA database. Notably, the correct benchmark allele was always included in the candidate set.

Consistent results were observed across sequencing platforms with different read-length distributions and error profiles, highlighting the robustness of QHap-derived haplotype information for downstream genomic analyses.

\begin{table*}[h!]
\centering
\captionsetup{justification=raggedright, singlelinecheck=false}
\caption{Cross-platform HLA typing results for the HG002 sample}
\resizebox{\linewidth}{!}{
    \begin{tabular}{lllll}
    \toprule
    HLA gene & Reference genotype & CycloneSEQ & HiFi & ONT \\
    \midrule
    A & A*26:01:01:01 & A*26:01:01 & A*26:01:01 & A*26:01:01 \\
    A & A*01:01:01:01 & \textbf{A*01:04:01:01N / A*01:01:01} & A*01:01:01 & A*01:01:01 \\
    B & B*38:01:01 & B*38:01:01 & B*38:01:01 & B*38:01:01 \\
    B & B*35:08:01 & B*35:08:01 & B*35:08:01 & B*35:08:01 \\
    C & C*12:03:01:01 & C*12:03:01 & C*12:03:01 & C*12:03:01 \\
    C & C*04:01:01:06 & C*04:01:01 & C*04:01:01 & C*04:01:01 \\
    DPA1 & DPA1*01:03:01 & DPA1*01:03:01 & DPA1*01:03:01 & DPA1*01:03:01 \\
    DPB1 & DPB1*04:01:01:01 & DPB1*04:01:01 & DPB1*04:01:01 & DPB1*04:01:01 \\
    DQA1 & DQA1*03:01:01 & DQA1*03:01:01 & DQA1*03:01:01 & DQA1*03:01:01 \\
    DQA1 & DQA1*01:05:01 & DQA1*01:05:01 & DQA1*01:05:01 & DQA1*01:05:01 \\
    DQB1 & DQB1*03:02:01 & \textbf{DQB1*03:02:01 / DQB1*03:02:24} & DQB1*03:02:01 & DQB1*03:02:01 \\
    DQB1 & DQB1*05:01:01:02 & DQB1*05:01:01 & DQB1*05:01:01 & DQB1*05:01:01 \\
    DRB1 & DRB1*04:02:01 & DRB1*04:02:01 & DRB1*04:02:01 & DRB1*04:02:01 \\
    DRB1 & DRB1*10:01:01 & DRB1*10:01:01 & DRB1*10:01:01 & DRB1*10:01:01 \\
    DRB4 & DRB4*01:03:01:01 & DRB4*01:03:01 & DRB4*01:03:01 & DRB4*01:03:01 \\
    G & G*01:06:01 & G*01:06:01 & G*01:06:01 & G*01:06:01 \\
    G & G*01:01:02 & G*01:01:02 & G*01:01:02 & G*01:01:02 \\
    \bottomrule
    \end{tabular}
}
\label{table:3}
\end{table*}

\section*{Discussion}

QHap reformulates haplotype phasing as a Max-Cut problem. This formulation enables the deployment of quantum-annealing-inspired algorithms, achieving substantial computational acceleration without compromising accuracy. Among these algorithms, bSB proves particularly effective due to its stable convergence across heterogeneous graph structures and natural compatibility with GPU parallelism. Our dual-strategy architecture accommodates the inherent tension between graph complexity and problem scale in long-read phasing: the read-based method captures direct fragment relationships for regional precision, while the SNP-based method reduces graph dimensionality for chromosome-scale tractability.

Current limitations, notably the elevated hamming error on chromosome-scale CycloneSEQ data, reflect the challenge of achieving global optimality within practical iteration limits as graph complexity increases. Chromosome-scale phasing inherently involves SNP-based formulations approaching \(10^{5}\) variables, necessitating further breakthroughs in quantum-inspired algorithm scalability. Alternatively, the read-based method offers a natural extension pathway. As sequencing technologies advance toward longer reads with higher accuracy, improved data quality will enable the construction of high-quality, contiguous unitigs in genomic regions of low repetitive content. This development will enable quantum-inspired phasing to systematically resolve complex repetitive regions, ultimately advancing the field toward chromosome-scale haplotype-resolved and telomere-to-telomere (T2T) assemblies.

Beyond immediate methodological advances, quantum-inspired algorithms represent a strategically valuable bridge toward practical quantum optimization techniques. The rapid development of distributed variational quantum algorithms is accelerating the deployment of quantum optimization in real-world applications. Critically, solutions obtained from quantum-inspired methods can serve as high-quality initial states for quantum circuits, enabling hybrid classical-quantum workflows where subsequent quantum refinement enhances convergence toward global optima. As both sequencing technologies and quantum hardware continue to mature, frameworks such as QHap will become increasingly essential for population-scale haplotype analysis, suggesting that quantum-inspired optimization may become a useful component of future large-scale phasing pipelines.

\section*{Methods}
\subsection*{QHap input preprocessing}

QHap requires a coordinate-sorted BAM file and a corresponding VCF file as input. To ensure high-fidelity allele assignments, QHap implements a rigorous quality-aware preprocessing pipeline. For each read covering heterozygous SNPs, the algorithm constructs a reference-to-read coordinate mapping through CIGAR string parsing, enabling the precise localization of variant positions within the read sequence. Allele determination requires dual thresholds: a minimum base quality (default Phred score $\geq$13) and mapping quality (default MAPQ $\geq$20). These thresholds filter out low-confidence observations that could introduce systematic errors into downstream phasing. For standard short-read and long-read data, reads are processed independently with variant-allele pairs aggregated into haplotype-informative fragments. When processing Pore-C data, QHap recognizes that multiple sub-alignments originating from the same chromatin complex share a common read identifier prefix; these are merged into unified fragments that preserve long-range haplotype connectivity across megabase-scale genomic distances. This fragment-merging strategy substantially increases the effective linking power of proximity ligation reads compared to treating each alignment independently. The preprocessed fragments, each encoding a sparse vector of phased allele observations with associated quality scores, serve as the primary input to the subsequent graph-based haplotype inference module.

\subsection*{Read-based method implementation}

\subsubsection*{Haplotype initialization}

The algorithm begins by establishing an initial haplotype configuration that serves as a reference framework for subsequent optimization. Given a set of heterozygous SNP positions $P = \{p_1, p_2, \ldots, p_n\}$ extracted from the input VCF file, we construct two complementary haplotype sequences $\mathrm{H1}$ and $\mathrm{H2}$. For each position $p_j$, the reference allele $r_j$ and alternative allele $a_j$ are randomly assigned to either $\mathrm{H1}$ or $\mathrm{H2}$. This initialization establishes a coordinate system for the ternary encoding in subsequent matrix construction, where read alleles are mapped relative to this initial assignment.

\subsubsection*{Matrix construction}

We construct a fragment-SNP matrix $M \in \{-1, 0, 1\}^{m \times n}$ from sequencing read data, where $m$ denotes the number of sequencing fragments and $n$ represents the number of heterozygous SNP positions. For each fragment $f_i$ with observed allele $b_{ij}$ at position $p_j$, the matrix entry is assigned as:
\begin{equation}
M_{ij} = 
\begin{cases}
+1 & \text{if } b_{ij} = \mathrm{H1}[j], \\
-1 & \text{if } b_{ij} = \mathrm{H2}[j], \\
0 & \text{otherwise.}
\end{cases}
\end{equation}

This ternary encoding transforms raw sequencing observations into a compact representation that captures haplotype concordance relationships. The method relies on the abundance of read evidence rather than individual quality metrics to resolve phasing ambiguities, treating all observations with equal reliability. The resulting sparse matrix structure efficiently accommodates long-read sequencing data, where fragments typically cover only a subset of the total SNP positions.

\subsubsection*{Max-Cut graph construction}

Building upon the ternary matrix $M$ constructed in the previous step, we transform the fragment-SNP relationships into a graph-based optimization framework. We construct a weighted undirected graph $G = (V, E)$ where vertices represent sequencing fragments and edges encode pairwise discordance relationships. This formulation transforms the haplotype phasing problem into a graph bipartition problem, where the objective is to separate fragments originating from different parental chromosomes.

For efficient edge weight computation, we first determine the column span for each SNP position, defined as the range of fragment indices containing non-zero observations at that position. Edge weights are computed by iterating through each column and examining all fragment pairs within its span. For fragments $f_a$ and $f_b$ both covering position $p_j$, a conflict is recorded when their ternary values differ while both are non-zero:
\begin{equation}
\delta_{ab}^{(j)} = \mathbf{1}[M_{aj} \cdot M_{bj} = -1],
\end{equation}
where $\mathbf{1}[\cdot]$ denotes the indicator function. The total edge weight between fragments $f_a$ and $f_b$ aggregates conflicts across all shared positions, which takes non-negative integer values:

\begin{equation}
w_{ab} = \sum_{j=1}^{n} \delta_{ab}^{(j)}.
\end{equation}

This conflict-counting approach provides a direct measure of haplotype inconsistency: fragments originating from the same haplotype should exhibit identical allelic patterns at shared positions (yielding $w_{ab}\approx 0$), while fragments from opposite haplotypes should display systematic differences (yielding large $w_{ab}$). The resulting positive edge weights naturally align with the Max-Cut formulation, where maximizing the total weight of cut edges corresponds to partitioning fragments that are most allele-discordant into opposite subsets.

To reduce computational complexity, the graph is decomposed into connected components using breadth-first search traversal. Each component represents an independent phasing subproblem that can be solved separately. Single-fragment components, which lack sufficient linkage information for phasing, are excluded from subsequent processing.

\subsubsection*{SB algorithm solution}

Each connected component is processed independently using the bSB algorithm on a GPU. Since the bSB algorithm minimizes an Ising-type Hamiltonian, we first define the coupling matrix $J$ based on the conflict-based adjacency matrix $A$ by:
\begin{equation}
J_{ab} = -A_{ab},
\end{equation}
where the sign inversion ensures that minimizing the Ising Hamiltonian is equivalent to maximizing the graph cut.
The corresponding Ising model (energy function) is given by:
\begin{equation}
\min E(\boldsymbol{s}) = -\frac{1}{2} \boldsymbol{s}^\mathrm{T} J \boldsymbol{s},
\end{equation}
where $\boldsymbol{s} \in \{-1, +1\}^{|V|}$ denotes the binary partition vector of fragments. This formulation is equivalent to the standard Max-Cut objective:
\begin{equation}
\max \frac{1}{4} \sum_{a,b} A_{ab} \left(1 - s_a s_b\right), \quad s_a \in \{-1, +1\}.
\end{equation}

The bSB algorithm operates through an iterative dynamics-based optimization process that evolves continuous spin and momentum variables. Through an adiabatic schedule, the algorithm progressively drives these variables toward binary values $\{-1, +1\}$, with boundary constraints and momentum resets ensuring numerical stability. Multiple initial configurations are sampled in parallel, and the final binary assignment is obtained via $s_a = \mathrm{sgn}(x_a)$, where $x_a$ represents the converged spin variable for fragment $f_a$ \cite{r18, tao2026tabu}.

The algorithm outputs a binary classification of fragments into two subsets $S_1$ and $S_2$, corresponding to the two parental haplotypes. The maximum cut value serves as a quality metric, with higher values indicating more accurate separation between haplotype-specific fragment groups.

\subsubsection*{Consistency processing}

Following fragment classification, a voting-based consensus procedure determines the final allele assignment at each SNP position. For each position $p_j$, we compute the supporting counts from both fragment subsets:
\begin{align}
c_1^{(+)} &= \sum_{i \in S_1} \mathbf{1}[M_{ij} = +1], & c_1^{(-)} &= \sum_{i \in S_1} \mathbf{1}[M_{ij} = -1], \\
c_2^{(+)} &= \sum_{i \in S_2} \mathbf{1}[M_{ij} = +1], & c_2^{(-)} &= \sum_{i \in S_2} \mathbf{1}[M_{ij} = -1],
\end{align}
where $c_l^{(+)}$ and $c_l^{(-)}$ denote the number of fragments in subset $S_l$ supporting the $\mathrm{H1}$ and $\mathrm{H2}$ alleles, respectively.

The final haplotype assignment employs a hierarchical decision rule based on differential support. When both subsets show concordant majority assignments (both favoring $+1$ or both favoring $-1$), the subset with the larger support differential $|c_l^{(+)} - c_l^{(-)}|$ retains its majority allele assignment, while the other subset is assigned the complementary allele to enforce bipartite haplotype structure. When subsets show opposing majorities, the subset favoring $+1$ is assigned allele $\mathrm{H1}[j]$, and that favoring $-1$ is assigned allele $\mathrm{H2}[j]$.

Positions exhibiting ambiguous evidence are filtered based on two criteria: (1) positions with zero coverage in both subsets, and (2) positions where both subsets show equal support for both alleles, indicating insufficient discriminative information. This strategy ensures that only high-confidence phasing calls are retained in the final output.

\subsection*{SNP-based method implementation}

\subsubsection*{Matrix construction}

The SNP-based method extends the basic ternary framework by incorporating sequencing quality information. The initial ternary matrix $\mathbf{M}$ is constructed as in the read-based method, but each entry is subsequently weighted by a confidence score derived from Phred quality values extracted from the BAM file.

For each fragment $f_i$ at position $p_j$, we compute a confidence weight $\tilde{q}_{ij}$ based on the Phred quality score $Q_{ij}$:
\begin{equation}
    \tilde{q}_{ij} = 1 - 10^{-Q_{ij}/10}.
\end{equation}

This transformation maps quality scores to confidence probabilities, where higher Phred scores yield weights approaching 1. The final weighted matrix $\mathbf{W}$ is obtained as:
\begin{equation}
    W_{ij} = M_{ij} \cdot \tilde{q}_{ij}.
\end{equation}

This unified representation encodes both allelic assignment and measurement confidence. Rather than applying additional stringent quality thresholds that would discard potentially informative observations, this soft weighting scheme preserves linkage information from lower-quality reads while ensuring that high-confidence observations dominate the final phase inference. This approach is particularly valuable in regions with sparse coverage, where even moderate-quality reads contribute essential connectivity for haplotype phasing.

\subsubsection*{Max-Cut graph construction}

Building upon the quality-weighted matrix $\mathbf{W}$, we transform the phasing problem into a graph-based formulation. Unlike the read-based method where vertices represent sequencing fragments, the SNP-based method constructs a weighted undirected graph $G = (V, E)$ with vertices $V$ corresponding to SNP positions and edges $E$ connecting positions jointly covered by sequencing fragments. Edge weights are derived from a probabilistic model that quantifies the evidence for \textit{cis} (same haplotype) versus \textit{trans} (different haplotype) relationships.

For positions $p_u$ and $p_v$ both covered by fragment $f_i$, let $q_u = |W_{iu}|$ and $q_v = |W_{iv}|$ denote their respective confidence weights. The probability that both positions are correctly assigned to the same haplotype (concordant assignment) is:
\begin{equation}
    P_{\textit{cis}} = q_u q_v + (1 - q_u)(1 - q_v).
\end{equation}

Conversely, the probability that positions $p_u$ and $p_v$ are assigned to different haplotypes (discordant assignment) is:
\begin{equation}
    P_{\textit{trans}} = q_u (1 - q_v) + (1 - q_u) q_v.
\end{equation}

The log-likelihood ratio for each fragment observation is computed as:
\begin{equation}
    l_{uv}^{(i)} = \text{sgn}(W_{iu} \cdot W_{iv}) \cdot \log_{10} \frac{P_{\textit{cis}}}{P_{\textit{trans}}},
\end{equation}
where the sign function $\text{sgn}(\cdot)$ accounts for the observed allelic relationship. Edge weights are aggregated across all fragments and normalized by the number of pairwise linkages within each fragment minus one, preventing fragments spanning many positions from dominating the total weight; the resulting edge weights are mixed-sign floating-point values:
\begin{equation}
\label{eq:edge_weight_SNP}
    w_{uv} = -\sum_{i \in F_{uv}} \frac{l_{uv}^{(i)}}{|S_i| - 1},
\end{equation}
where $F_{uv}$ denotes the set of fragments covering both positions $p_u$ and $p_v$, and $|S_i|$ is the number of heterozygous positions covered by fragment $f_i$. The negation ensures that maximizing the cut corresponds to maximizing phasing consistency. Subsequently, connected components of the graph are identified, allowing for independent processing of disjoint genomic regions.

\subsubsection*{SB algorithm solution}

As in the read-based method, the SNP-based method employs the bSB algorithm to solve the Max-Cut problem on GPU. However, because the SNP-based graph encodes aggregated evidence from multiple fragments per edge rather than direct pairwise fragment conflicts, it incorporates an iterative refinement procedure to resolve the resulting more complex optimization landscape arising from mixed-sign floating-point edge weights.

After obtaining an initial partition with two subsets $(S_1, S_2)$, we identify the smaller subset and update a position-wise flip counter $C(p_u)$ for positions in this subset. The graph weights are then updated by inverting the sign of edges crossing the current partition boundary:
\begin{equation}
    w'_{uv} = 
    \begin{cases}
        -w_{uv} & \text{if } s_u \neq s_v, \\
        w_{uv} & \text{otherwise},
    \end{cases}
\end{equation}
where $s_u \in \{-1, +1\}$ denotes the partition assignment of SNP position $p_u$ obtained from the current bSB solution. This weight inversion mechanism progressively refines the phasing by accumulating evidence across iterations. Let $S_{\min}^{(t)}$ denote the smaller subset at iteration $t$; the flip counter accumulates as:
\begin{equation}
    C(p_u) = \sum_{t=1}^{T} \mathbf{1}[p_u \in S_{\min}^{(t)}],
\end{equation}
where $T$ is the total number of iterations. The iteration terminates when the cut value becomes non-positive or the partition converges to a state where all vertices belong to a single subset, indicating that the current haplotype assignment is at least locally optimal.

The final haplotype assignment is determined by the parity of flip counts. Since each increment of $C(p_u)$ corresponds to one phase inversion relative to the initial assignment, the reconstructed haplotype pair $(\mathrm{H1}', \mathrm{H2}')$ is derived from the initial pair $(\mathrm{H1}, \mathrm{H2})$ as:
\begin{equation}
\hspace{-1pt}
    (\mathrm{H1}'[u], \mathrm{H2}'[u]) = 
    \begin{cases}
        (\mathrm{H1}[u], \mathrm{H2}[u]) & C(p_u) \bmod 2 = 0, \\
        (\mathrm{H2}[u], \mathrm{H1}[u]) & C(p_u) \bmod 2 = 1.
    \end{cases}
\end{equation}

For challenging instances exhibiting persistent negative cut values, we employ an adaptive parameter strategy that introduces a deformation term $h$ to escape local minima, with iteration counts dynamically adjusted based on graph size to balance solution quality and computational efficiency. The entire pipeline leverages GPU acceleration through PyTorch sparse tensor operations, achieving substantial speedup over CPU-based implementations for large-scale genomic datasets, as demonstrated in Figure \ref{fig:3}.

\subsection*{Pore-C data support}

\begin{figure*}[h!]
    \centering
    \includegraphics[width=\textwidth,keepaspectratio]{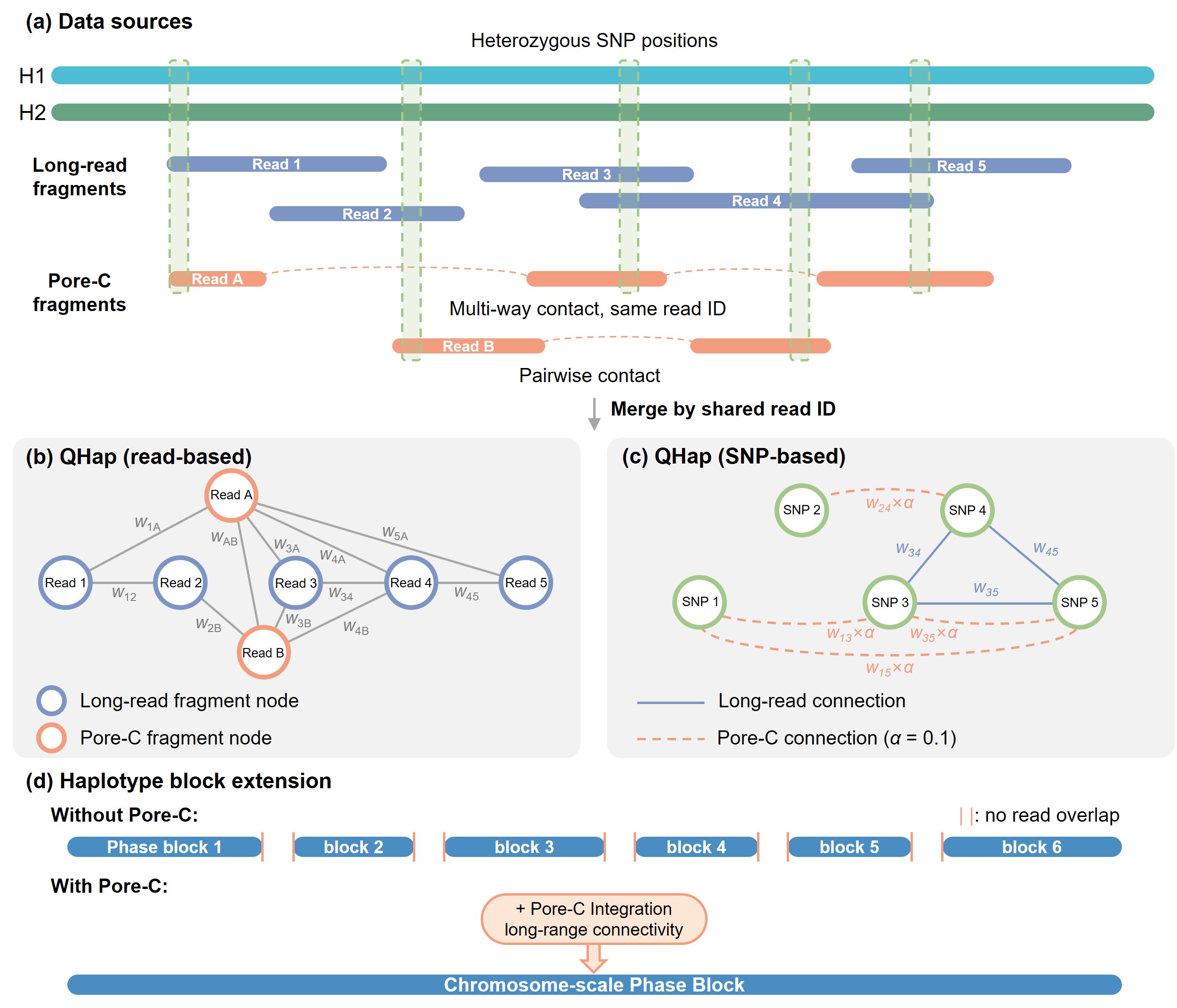}
    \caption{\textbf{Schematic overview of Pore-C data integration in the
    QHap framework.}
    Conventional long-read sequencing generates continuous fragments
    covering local genomic regions, while Pore-C produces multi-way contact fragments that link distant loci
    via chromatin looping; sub-alignments sharing a common read identifier
    are merged into unified fragments~(\textbf{a}). These two data
    sources are integrated through strategy-specific graph construction
    approaches. In the read-based method~(\textbf{b}), all fragment
    nodes from both long-read and Pore-C data are incorporated into the
    graph with uniform edge weighting based on allelic conflict counts,
    relying on read abundance for phase resolution. In the SNP-based
    method~(\textbf{c}), edges derived from long-read data carry full
    weight, whereas Pore-C-derived edges are down-weighted by a factor of
    $\alpha = 0.1$ to mitigate elevated error rates while preserving
    long-range connectivity. Through this balanced integration, Pore-C
    data bridges gaps between isolated phase blocks caused by homozygous
    regions or insufficient read overlap~(\textbf{d}), enabling
    near-chromosome-scale haplotype reconstruction that combines extended
    contiguity with high local phasing precision.}
    \label{fig:6}
\end{figure*}

Both the read-based and SNP-based methods support Pore-C data
integration to extend phasing contiguity, as illustrated in
Figure~\ref{fig:6}. The approach first integrates variant calls and
fragment data from both conventional long-read and Pore-C sequencing
platforms, resolving allelic conflicts at overlapping genomic positions
by prioritizing variants with superior quality metrics.

For the read-based method, Pore-C fragments are incorporated into the
fragment graph without differential weighting, consistent with the
method's reliance on read abundance rather than individual quality
metrics for phase resolution (Figure~\ref{fig:6}b).

For the SNP-based method, to harness the ultra-long-range connectivity
of Pore-C data while mitigating its elevated error rate, the
contributions of Pore-C-derived fragments to the final edge weights
$w_{uv}$ (Eq.~\ref{eq:edge_weight_SNP}) are scaled by a factor
$\alpha$ (default 0.1) during graph construction
(Figure~\ref{fig:6}c). This weighting scheme strategically exploits
the complementary strengths of both data types: Pore-C reads establish
critical long-range phasing scaffolds spanning megabase-scale haplotype
blocks, while local high-confidence phase calls are predominantly
informed by higher-accuracy long-read data.

This balanced integration framework enables the reconstruction of
chromosome-scale haplotypes with both extended continuity and high
local precision (Figure~\ref{fig:6}d).

\subsection*{Algorithm of HLA typing}

\subsubsection*{Variant calling and haplotype phasing}

Raw sequencing reads were aligned to the CHM13v2 human reference genome using Minimap2. Reads mapped to HLA loci were extracted to generate HLA-specific BAM files. Variants within these regions were identified using DeepVariant to detect SNPs and small insertions or deletions. Heterozygous SNPs were phased using the strategy described in this study. The resulting haplotypes were assigned to sequencing reads using the haplotag function in WhatsHap, and haplotype-specific read sets were subsequently extracted using SAMtools.

\subsubsection*{Read binning}

The IMGT/HLA database containing 6,172 alleles from 39 HLA genes and pseudogenes was used as the reference allele set. For each haplotype-specific read set, reads mapped to the HLA region were compared with allele sequences in the IMGT/HLA database and assigned to the best-matched allele based on the minimal number of mismatches. Reads were then grouped according to the HLA gene to which the matched allele belongs, generating candidate allele sets and corresponding read groups for each locus.

\subsubsection*{HLA allele assignment}

HLA typing was performed independently for each gene and haplotype. Reads assigned to each locus were used to reconstruct a haplotype-specific consensus sequence by selecting the most supported nucleotide at each position. The consensus sequence was subsequently aligned against allele sequences of the corresponding gene in the IMGT/HLA database using the Wavefront Alignment Algorithm. The allele with the highest sequence similarity was selected as the typing result. Finally, typing results from both haplotypes were combined to determine the HLA genotype of the sample.

\subsection*{Evaluation metrics of haplotype phasing}

To comprehensively assess the accuracy and efficiency of haplotype phasing methods, we employ six evaluation metrics encompassing phasing quality, structural characteristics, completeness, and computational requirements.

\textbf{Switch error rate.} This metric quantifies phase inconsistencies within reconstructed haplotypes. It measures the frequency of adjacent heterozygous SNP pairs with incorrect relative phase orientation compared to the reference. Formally, the SE is computed as the ratio of switch positions to $n-1$ consecutive SNP pairs for a sequence of $n$ SNPs. For instance, consider an inferred haplotype sequence ``01110'' against a reference sequence ``00000'': here, phase switches occur at two positions, yielding $\text{SE} = 2/(5-1) = 50\%$.

\textbf{Hamming error rate.} This measures the proportion of incorrectly assigned alleles across all phased heterozygous sites when comparing the phased haplotype against either the paternal or maternal reference haplotype. Using the previous example, where three alleles differ between ``01110'' and ``00000'', the HE equals $3/5 = 60\%$. This metric provides a site-level accuracy assessment independent of local phase consistency.

\textbf{Haplotype N50.} To characterize the contiguity of phased regions, we calculate the N50 statistic of haplotype blocks, where each block represents a continuous sequence of phased variants. The N50 value is determined by rank-ordering blocks in descending size and identifying the length at which the cumulative sum reaches 50\% of the total phased genomic span.

\textbf{Phasing completeness.} This metric reflects the coverage of successful phasing, defined as the proportion of phased heterozygous sites relative to all identifiable heterozygous loci in the input dataset. It indicates the algorithm's ability to assign phase information across available variant positions.

\textbf{SNP completeness.} This metric represents the fraction of heterozygous SNPs retained in the final phased output compared to the total number of true heterozygous variants present in the reference dataset. Unlike phasing completeness, which measures coverage among detected variants, SNP completeness can be used to evaluate the algorithm's filtering strategy or its ability to balance phasing accuracy against completeness. 

\textbf{Running time.} We record the total execution time encompassing all processing stages, including data loading, preprocessing operations, and the core phasing algorithm execution, providing a practical measure of computational scalability.

\section*{Code availability}
All numerical simulations presented in this work were implemented using the MindSpore Quantum framework \cite{xu2024mindspore}, which provides a powerful environment for quantum computation simulation. The source code and data required to reproduce the numerical results reported in this paper are publicly available at \url{https://github.com/Zr1311/QHap.git}.

\section*{Acknowledgements}

This study is supported by the Shenzhen International Quantum Academy (Grant No. SIQA2025KFKT06), the CPS-Yangtze Delta Region Industrial Innovation Center of Quantum and Information Technology-MindSpore Quantum Open Fund, and the Shenzhen Science and Technology Program (Grant No. SYSPG20241211173852024).
  
\bibliography{sample}          

\end{document}